\documentclass[aps,10pt,pra,twocolumn,amsmath,amssymb,groupedaddress,floatfix,showpacs]{revtex4-1} 
\usepackage{graphicx}  

\newcommand{\Hil}{\mathcal{H}}
\newcommand{\Sys}{\mathcal{S}}
\newcommand{\E}{\mathcal{E}}

\newcommand{\ket}[1]{\vert #1 \rangle}
\newcommand{\bra}[1]{\langle #1 \vert}
\newcommand{\op}[2]{\vert#1\rangle\langle#2\vert}

\newcommand{\abs}[1]{\vert #1 \vert}
\newcommand{\re}[1]{\text{Re}\left\{#1\right\}}
\newcommand{\im}[1]{\text{Im}\left\{#1\right\}}
\newcommand{\tr}[2][s]{\text{tr}_{#1}\{#2\}}
\newcommand{\T}{\mathcal{T}}
\newcommand{\J}{\mathcal{J}}

\newcommand{\ddt}[1]{\frac{\text{d}}{\text{d}#1}}
\newcommand{\eff}{\text{eff}}
\newcommand{\N}{\mathcal{N}}

\begin{document}

\author{Kimmo Luoma}

\email{ktluom@utu.fi}

\author{Kalle-Antti Suominen}

\author{Jyrki Piilo}
\affiliation{Turku Centre for Quantum Physics, Department of Physics and 
Astronomy, University of Turku, FI-20014, Turun Yliopisto, Finland}

\date{\today}

\title{Connecting two jumplike unravelings for non-Markovian 
open quantum systems}


\begin{abstract}
The development and use of Monte Carlo algorithms plays a visible role in the 
study of non-Markovian quantum dynamics due to the provided insight and 
powerful numerical methods for solving the system dynamics. In the Markovian 
case, the connections between the various types of methods are fairly 
well-understood while for non-Markovian case there has so far been only a 
few studies. We focus here on two jumplike unravelings of non-Markovian 
dynamics, the non-Markovian quantum jump (NMQJ) method and the property state 
method by Gambetta, Askerud, and Wiseman (GAW). The results for simple quantum 
optical 
systems illustrate the connections between the realizations of the two methods 
and also highlight how the probability currents between the system and 
environment, or between the property states of the total system, associate 
to the decay rates of time-local master equations, and consequently to the 
jump rates of the NMQJ method.
\end{abstract}

\maketitle

\section{Introduction}
The theory of open quantum systems deals with the dynamics of the reduced 
system which is coupled to its
environment~\cite{Breuer2007}. This leads often to 
decoherence and the loss of 
quantum properties~\cite{PhysRevD.24.1516,PhysRevD.26.1862,PalmaSuominenEkert,Dechoherence_Classical_Quantum2003}, though there also exists schemes to 
exploit system-reservoir interaction for quantum engineering~\cite{Myatt2000,Diehl2008,Verstraete2009}.
Recently, non-Markovian dynamics, where memory effects play a crucial role, 
has become under very active research~\cite{PhysRevLett.88.170407,PhysRevA.70.010304,Piilo2008,Piilo2009,BreuerPiilo2009,PhysRevLett.102.100402,PhysRevLett.103.210401,PhysRevLett.104.070406,PhysRevA.81.052103,PhysRevA.82.052111,PhysRevA.83.032103,PhysRevA.83.042321,2011arXiv1104.5263Z}. On one hand this is due to the fact that the 
fundamental understanding of non-Markovianity is still missing and on the 
other hand non-Markovianity may be useful for various quantum information or 
quantum engineering tasks~\cite{2011arXiv1103.1219C,PhysRevA.83.042321}.

The solving of non-Markovian dynamics is often a challenging task and there 
exists a large number of both analytical methods~\cite{Breuer2007,PTP.20.948,zwanzig:1338,Breuer200136,PhysRevE.73.016139} and numerical Monte Carlo 
algorithms for this purpose~\cite{PhysRevA.50.3650,Garraway1996,Diosi1997569,PhysRevLett.82.1801,PhysRevA.59.1633,PhysRevA.59.2306,pseudomodes,PhysRevA.64.053813,PhysRevA.66.012108,PhysRevA.69.052104,PhysRevA.70.012106,BreuerPiilo2009,Piilo2009,Piilo2008}.
Roughly speaking, the Monte Carlo methods can be divided into discontinuous 
jumplike unravelings or continuous diffusion type unravelings. For the 
Markovian case without memory effects, the connections between the methods are 
fairly well understood~\cite{Breuer2007,Zoller:92,PhysRevLett.68.580,PhysRevA.46.4363,Gisin:92,Molmer:93,Carmichael2008} while the same can not be said of the 
non-Markovian methods despite of a few early studies \cite{PhysRevA.66.012108,PhysRevA.80.012104}. 

We focus here on two jumplike unravelings and illustrate their connections by 
studying simple quantum optical systems. The method by 
Gambetta, Askerud, and Wiseman (GAW) is based on the generating stochastic 
realizations for the total system state vectors and monitoring the random 
jumps between the property states of the total 
system~\cite{PhysRevA.69.052104}. On the other hand, the recently developed 
non-Markovian quantum jump (NMQJ) method generates jumplike realizations for 
the state vectors within the Hilbert space of the reduced system. 

We show here that there is an inherent connection between the reduced system 
part of the GAW realizations and the NMQJ realizations. 
Moreover, we also 
study how the probability currents between the property states are associated 
to the decay rates of the time-local master equations, and how the jump rates 
between the GAW and NMQJ methods are connected. 
We stress that the NMQJ method can be currently used for systems for which 
time local non-Markovian master equation can be derived whereas GAW 
method has greater 
generality.
Our results provide new 
insight for non-Markovian dynamics in terms of the information flow between 
the system and the reservoir, and hopefully stimulates further studies of 
connections between Monte Carlo methods for non-Markovian dynamics.

The paper is organized in the following way. In Section~\ref{sec:GAW} we 
describe the basic ingredients of the GAW method and in Sec.~\ref{sec:NMQJ} 
of the NMQJ method. By studying simple quantum optical systems, 
in Sec.~\ref{sec:Conn} we show how the methods are connected and finally, 
Sec.~\ref{sec:Conc} concludes the paper.

\section{Unraveling in the total system space: GAW method}\label{sec:GAW}

The method by Gambetta, Askerud and Wiseman (GAW) is based on generating 
piecewise deterministic realizations, or jumplike unraveling within the 
Hilbert space of the total system, describing the discontinuous transitions 
between the property states of the system. We give here the basic ingredients 
of the method suitable for undriven quantum optical systems with 
spectral mode unraveling. We note that GAW method can 
also be applied to driven systems and with temporal mode unraveling.  
More details can be found from the 
Refs.~\cite{PhysRevA.69.052104,FoundPhys.34.419,PhysRevA68.062104}. 

We focus on the dynamics of simple undriven quantum optical systems, 
e.g., two-level 
and V-systems which are coupled to a continuum of electromagnetic field modes 
at zero temperature. The dynamics of the state vector of the system and the 
environment in $\Hil_\Sys\otimes\Hil_\E$, where $\Hil_\Sys$ and $\Hil_\E$ are 
the Hilbert spaces of the system and the environment respectively, is given 
by the Schr\"odinger
equation
\begin{align}\label{eq:Schrodinger}
\frac{\text{d}}{\text{d}t}\ket{\Psi(t)}=&-iH\ket{\Psi(t)}.
\end{align}
Here we have set $\hbar=1$.
The Hamiltonian $H=H_\Sys + H_\E + H_{\Sys\E}$ includes the free evolution of the system $H_\Sys$ and the environment $H_\E$, and the system-environment 
interaction $H_{\Sys\E}$.
The free evolution of the $n$-level system is governed by 
$H_\Sys=\sum_{k=1}^n\omega_k\op{k}{k}$, where 
$\ket{k}$ are the energy eigenstates of the system and $\omega_k$ the corresponding energies.
The free evolution of the $N$-mode environment is given by
$H_\E=\sum_{j=1}^N\nu_ja_j^\dagger a_j$, where the operators 
$a_j$($a_j^\dagger$) are the annihilation (creation) operators
for $j$th mode of the environment. 
For simplicity  we focus on system-environment interactions 
that under the rotating wave approximation (RWA) 
include only transition from the excited states to the 
unique ground state without any cascade structure.
The general form of such interaction is
\begin{align}
H_{\Sys\E}=i\sum_{k > 1}^n\sum_{j=1}^N(g_k\op{1}{k}a_j^\dagger-
g_k^*\op{k}{1}a_j).
\end{align}
From now on we will work in the interaction picture
$H\rightarrow H_I(t)=e^{i(H_\Sys +H_\E)t}He^{-i(H_\Sys +H_\E)t}$, 
where the dynamics is given by Eq.~\eqref{eq:Schrodinger}
with the Hamiltonian
\begin{align}\label{eq:H_I}
H_I(t)=  i\sum_{k > 1}^n\sum_{j=1}^N(g_k\op{1}{k}a_j^\dagger e^{-i\Omega_{j,k}t}-
g_k^*\op{k}{1}a_j e^{i\Omega_{j,k}t}),
\end{align}
where $\Omega_{j,k}=\nu_j-\omega_k$ and $\omega_k$ is the 
energy difference of the ground state (labeled with index 1) and
$k$th excited state of the system. Here $g_k$  is a frequency-dependent coupling constant.

The total system state vector $\ket{\Psi(t)}$ evolves according to the
Eq.~\eqref{eq:Schrodinger} with the Hamiltonian \eqref{eq:H_I}.
Let us define a projective operator valued measure (POVM) as
\begin{align}\label{eq:ProjOpMes}
\pi_{m_N}&=I_\Sys\otimes_{j=1}^N\op{n_j}{n_j}\notag \\
&=I_\Sys\otimes\op{m_N}{m_N},
\end{align} 
where $m_N$ is a shorthand notation for arbitrary
photon number configuration of the $N$ environmental modes. In the 
systems we study we can 
have at maximum one excitation in the environment. However, GAW method 
in general is not limited only to one excitation~\cite{PhysRevA.69.052104}. 
We can now define property states of the total system  which are 
conditioned on some particular photon number 
configuration of the environment. These are
\begin{align}\label{eq:PropertyState}
\ket{\Psi_{m_N}}&=\pi_{m_N}\ket{\Psi(t)}/\sqrt{\N_{m_N}}\notag\\
&=\frac{1}{\sqrt{\N_{m_N}}}\ket{\phi_{m_N}(t)}_\Sys\otimes\ket{m_N}_\E,
\end{align}
where $\N_{m_N}$ is a normalization factor. We denote
the unnormalized property states with $\ket{\tilde{\Psi}_{m_N}(t)}$.
From now on we drop the subscript $N$ for notational convenience and the simple
index $m$ refers to a particular 
configuration of $N$ environmental modes.

The GAW method is a piecewise deterministic process (PDP) where the jumps take place between the 
different property states $\ket{\Psi_{m}(t)}$~\cite{Bell1984,PhysRevA.69.052104}. 
Let us define $P(m,t)$ as the 
probability for the total system to be in
the state $\ket{\Psi_{m}(t)}$ at time $t$ and 
write the following equation 
of motion for $P(m,t)$:
\begin{align}\label{eq:GAWMaster}
  \ddt{t}P(m,t)=\sum_{k}J_{m,k}(t),
\end{align}
where $J_{m,k}(t)$ 
is the probability current 
from $\ket{\Psi_{k}(t)}$ to $\ket{\Psi_{m}(t)}$, when $J_{m,k}(t)>0$, and 
when $J_{m,k}(t)<0$ it is the probability current from from $\ket{\Psi_{m}(t)}$ to $\ket{\Psi_{k}(t)}$, i.e.\ to the opposite direction. 
We can 
define this in the following way
\begin{align}
J_{m,k}(t)=T_{m,k}(t)P(k,t)-T_{k,m}(t)P(m,t),
\end{align}
where $T_{m,k}(t)$ is the transition rate 
from $\ket{\Psi_{k}(t)}$ to
$\ket{\Psi_{m}(t)}$. From this definition it is clear 
that $J_{m,k}(t)=-J_{k,m}(t)$. 
Given $J_{m,k}(t)$ and $P(m,t)$ there are many possible transition rates
satisfying Eq.~\eqref{eq:GAWMaster}. One possibility is to use the following one~\cite{PhysRevA.69.052104,Bell1984}:

\noindent When $J_{m,k}(t)\geq 0$,
\begin{align}\label{eq:GAWTPos}
  T_{m,k}(t)&=\frac{J_{m,k}(t)}{P(k,t)},\notag\\
  T_{k,m}(t)&=0,
\end{align}
and when  $J_{m,k}(t)<0$,
\begin{align}\label{eq:GAWTNeg}
  T_{m,k}(t)&=0,\notag\\
  T_{k,m}(t)&=-\frac{J_{m,k}(t)}{P(m,t)}.
\end{align}
Since we know the total wave function of the system and the environment 
$\ket{\Psi(t)}$, the probability of a given property state  $\ket{\Psi_{m}(t)}$ is
$P(m,t)=\bra{\Psi(t)}\pi_{m}\ket{\Psi(t)}$.
Using $P(m,t)$ and Eq.~\eqref{eq:Schrodinger} with the Hamiltonian in Eq.~\eqref{eq:H_I} we obtain
\begin{align}\label{eq:J}
J_{m,k}(t)=2\im{\bra{\Psi(t)}\pi_{m}H_I(t)\pi_{k}\ket{\Psi(t)}}.
\end{align}

The method for generating the realizations of the process begins with
solving the total system Schr\"odinger equation followed by the calculation of the quantities
$J_{m,k}(t)$, $T_{m,k}(t)$ and $P(k,t)$. For example, the probability to 
have a jump between the property states of the total system  $\ket{\Psi_{k}(t)}\rightarrow\ket{\Psi_{m}(t)}$,
when $J_{m,k}(t)>0$, between $[t,t+\delta t]$ is 
$\delta t T_{m,k}(t)$. 
Then using random numbers we 
can decide whether a jump takes place or not.
From Eqs.~\eqref{eq:GAWTPos} and~\eqref{eq:J} we see that the term
$\delta t T_{m,k}(t)$ 
includes the occupation probability of the source state
$k$ in the ensemble and the rate term from $k$ to $m$ and 
together they give the transition rate for a single trajectory.

After generation of the realizations, the 
state of the reduced system is
\begin{align}
\rho_\Sys(t)=&\tr[\E]{\sum_{m}w_{m}(t)\op{\Psi_{m}(t)}{\Psi_{m}(t)}}\notag\\
=&\sum_{m}w_{m}(t)\op{\phi_{m}(t)}{\phi_{m}(t)},
\end{align}
where $w_{m}(t)=\frac{\#(m)}{M}$ are the approximations 
for probabilities $P(m,t)$, $M$ is the size of the statistical ensemble and 
$\#(m)$ is the number of ensemble members in state $m$.

\section{Unraveling in the reduced system space: NMQJ method}\label{sec:NMQJ}

The non-Markovian quantum jump (NMQJ) method is constructed as a piecewise 
deterministic process in the Hilbert space of the system 
$\Hil_\Sys$~\cite{Piilo2009,Piilo2008,BreuerPiilo2009} and the key ingredient 
is the association of negative decay rates to reverse quantum jumps. The 
starting point is the time-local non-Markovian master equation, which 
can be derived, e.g. with time-convolutionless 
projection operator method (TCL)~\cite{Breuer2007}.
General form of such master equation (given here in the interaction picture) 
is
\begin{align}\label{eq:NMMaster}
\frac{\text{d}}{\text{d}t}\rho_\Sys(t)&=
-i[H_{\text{LS}}(t),\rho_\Sys(t)]\notag \\
&+\sum_j\Delta_j(t)
\bigg(C_j\rho_\Sys(t)C_j^\dagger-
\frac{1}{2}\{\rho_\Sys(t),C_j^\dagger C_j\}\bigg),
\end{align}
where $H_{LS}(t)=\frac{1}{2}\sum_jS_j(t)C_j^\dagger C_j$ 
is the Lamb shift Hamiltonian, $S_j(t)$ is 
the Lamb shift rate, $\Delta_j(t)$ is the decay rate 
which can take negative values and operator $C_j$ is the Lindblad or jump operator to channel $j$. 
The density matrix of the system at any point of time is decomposed as 
\begin{align}\label{eq:NMQJrho}
\rho_\Sys(t)=\sum_{i=1}^{M_{\text{eff}}}
P\left(\ket{\psi_i(t)},t\right)\op{\psi_i(t)}{\psi_i(t)},
\end{align}
where $M$ is the size of the statistical ensemble of the unraveling, $M_{\text{eff}}$ is
the dimension of the set of different states needed in the simulation 
(so called effective ensemble size), and $P(\ket{\psi_i(t)},t)$ is the probability
of finding state $\op{\psi_i(t)}{\psi_i(t)}$ in $\rho_\Sys(t)$.
The states in the ensemble 
evolve according to~\cite{Breuer2007,Molmer:93}
\begin{align}\label{eq:NMQJdet}
\frac{\text{d}}{\text{d}t}\ket{\psi_i(t)}&=-iH_\text{eff}(t)\ket{\psi_i(t)}
\notag\\
&=-i\left(H_{\text{LS}}(t)-i\frac{1}{2}\sum_j\Delta_j(t)C_j^\dagger C_j\right)\ket{\psi_i(t)}.
\end{align}

The rate of jumps during positive decay in channel $j$ from state 
$\ket{\psi_k(t)}$ to 
state $\ket{\psi_l(t)}$ with jump operator $C_j$ is 
\begin{align}\label{eq:NMQJRPos}
R_{lk}^j(t)=\Delta_j(t)\bra{\psi_k(t)}C_j^\dagger C_j\ket{\psi_k(t)}.
\end{align}
The corresponding quantum jump is given by
\begin{align}\label{eq:Jump}
\ket{\psi_k(t)}\rightarrow\ket{\psi_l(t)}=
\frac{C_j\ket{\psi_k(t)}}{\sqrt{\bra{\psi_k(t)}C_j^\dagger C_j\ket{\psi_k(t)}}}.
\end{align}  
The action of operator $C_j$ thus means that the  
state $\ket{\psi_k(t)}$ is destroyed and the state $\ket{\psi_l(t)}$ is 
created in the statistical ensemble.

During a negative decay probability period the jumps occur in reverse direction in the following sense:
\begin{align}\label{eq:JumpNeg}
\ket{\psi_k(t)}\leftarrow\ket{\psi_l(t)}=
\frac{C_j\ket{\psi_k(t)}}{\sqrt{\bra{\psi_k(t)}C_j^\dagger C_j\ket{\psi_k(t)}}}.
\end{align}  
The rate of these reverse jumps is obtained from 
\begin{align}\label{eq:NMQJRNeg}
R_{kl}^j(t)=-\frac{P(\ket{\psi_k(t)},t)}{P(\ket{\psi_l(t)},t)}
\Delta_j(t)\bra{\psi_k(t)}C_j^\dagger C_j\ket{\psi_k(t)}.
\end{align}

\section{Connection between the GAW and NMQJ unravelings}\label{sec:Conn}

To make a connection between the two unravelings, we are interested (i) whether the reduced system part of the total system property state realizations of the GAW method have similarities with the NMQJ realizations, and (ii) if the jumps within the two methods occur with the same rates. 
As we will show below, the answer for both of these questions is positive.

Comparing the rates, Eqs.~\eqref{eq:GAWTNeg} and~\eqref{eq:NMQJRNeg}, we note that 
the jump rates for the reverse probability flow and negative decay rates, $J_{m,k}<0$ and $\Delta_j(t)<0$
respectively, have similar structure. They both are inversely proportional to the probability to be in the source state of the jump. In the GAW method, the given property state is associated to specific mode to have the excitation (unless the environment is in the vacuum state). In the NMQJ realizations, we know whether the system or the environment has the excitation while in the latter case we do not know which specific mode has the excitation.

In order to reveal the detailed connection between the GAW and NMQJ methods,  let us define the 
following operators
\begin{align}\label{eq:CollProj}
\Pi_0=&I_\Sys\otimes\op{0_1,0_2,...,0_N}{0_1,0_2,...,0_N}=I_\Sys\otimes\op{0}{0},
\notag \\
\Pi_1=&\sum_{k=1}^NI_\Sys\otimes a_k^\dagger\op{0}{0}a_k=
\sum_{k=1}^NI_\Sys\otimes \op{1_k}{1_k}.
\end{align}
From Eq.~\eqref{eq:ProjOpMes} we see that $\Pi_0=\pi_{m=0\cdots0}$ and 
$\Pi_1=\sum_{k}\pi_{k}$, where $k=0\cdots1_k\cdots0$ (ie. $k$ labels all 
one mode configurations of the environment).
We can now ask what is the 
probability $P(0,t)$ to find zero photons at time $t$ in the environment?
This is given by 
\begin{align}\label{eq:P_0}
P(0,t)=\bra{\Psi(t)}\Pi_0\ket{\Psi(t)}. 
\end{align}
Similarly, the total probability $P(1,t)$ of having one photon in the environment, but not 
knowing in which mode, is  
\begin{align}\label{eq:P_1}
P(1,t)=\bra{\Psi(t)}\Pi_1\ket{\Psi(t)}. 
\end{align}
The connection between the GAW and the NMQJ methods is found by re-formulating the GAW method for the 
following combined property states:
\begin{align}\label{eq:ModPropertyState} 
\ket{\Psi_0(t)}=&\frac{1}{\sqrt{\N_0}}\Pi_0\ket{\Psi(t)},\notag \\ 
\ket{\Psi_1(t)}=&\frac{1}{\sqrt{\N_1}}\Pi_1\ket{\Psi(t)}. 
\end{align}
For this purpose, we must calculate the combined probability current
from the system to the environment. This is obtained by considering the total probability
current from the $N$-mode vacuum states to all $1_k$-states 
\begin{align}\label{eq:Jtot}
\J_{1,0}(t)=\sum_{k=1}^N J_{1_k,0}(t).
\end{align}
It can be easily shown that the combined probability current satisfies $\J_{1,0}(t)=-\J_{0,1}(t)$
and 
\begin{align}\label{eq:CombinedGAWMaster}
\ddt{t}P(1,t)&=\J_{1,0}(t),\notag \\
\ddt{t}P(0,t)&=-\J_{1,0}(t).
\end{align} 
We have 
$\sum_{k=1}^N\ddt{t}P(1_k,t)=\ddt{t}P(1,t)$ and the r.h.s.\ of both equations
follow from the definition of Eqs.~\eqref{eq:GAWMaster} and~\eqref{eq:Jtot}.
The transition rates have similar structure as in
Eqs.~\eqref{eq:GAWTPos} and \eqref{eq:GAWTNeg} 
but we must replace probability with combined probability and 
probability current with combined probability current.  

As we will show below for specific examples, the GAW transition rates defined with combined quantities correspond to the transition rates of the NMQJ method. Here, $\ket{\Psi_1(t)}$ and $\ket{\Psi_0(t)}$,  which are defined in the total system Hilbert space $\Hil_\Sys\otimes\Hil_\E$, are the possible values of the stochastic wave function of the combined GAW process. If the system part belonging to $\Hil_\Sys$ of the GAW stochastic wave function is in the same projective ray as the values of the stochastic wave function of the NMQJ method, we can conclude that both methods generate similar realizations for the reduced system. We will also see that the deterministic evolutions of the stochastic wave functions for both processes are identical.

This means that the PDPs of the two methods are the same in the following sense. The state space consists of the same set of projective rays in $\Hil_\Sys$, stochastic wave functions evolve similarly between random jumps in both processes, and random jumps in both processes take place between the same two projective rays in $\Hil_\Sys$ with equal rates. It is sufficient that the states belong to the same projective ray in $\Hil_\Sys$ since we are interested only in the dynamics of the reduced system. Moreover, the states in the same projective ray give equal contribution to the density matrix of the system since the complex phase of the state is not an observable.
 
The summing of the GAW probability currents means that we lose the information to which mode
the excitation from the system goes as the system decays. It is 
intuitive that the sum of the probability currents corresponds to 
the decay rate since the decay rate describes the total effect of the environment 
onto the system. However, it is important to note that for a given sign of the decay rate, there typically occurs probability flow components of the GAW 
realizations to both directions.

In examples below, we will set the frequency dependent couplings to be real-valued and
equal to $g_k=\sqrt{\text{d}\nu \rho_k(\nu_k)}$, where 
$\rho_k(\nu)=\frac{1}{2\pi}\frac{\gamma_0\lambda^2}{(\nu-\omega_c)^2+\lambda^2}$
is the spectral density, $\text{d}\nu$ is the mode spacing, $\lambda$ is the
spectral width, $\omega_c$ is the position of the peak in
frequency space, and $\gamma_0$ defines the height of the peak. These parameters 
are also related to the time scales involved in the dynamics. 
We have $\tau_\Sys \sim \gamma_0^{-1}$, which is the time scale of the 
reduced system evolution,  and $\tau_\E \sim \lambda^{-1}$ is 
the time scale of the environmental correlation functions.
We can also compare our discrete $N$-mode cases to 
the exact and numerical solutions obtained in the continuum limit 
$\sum_k\abs{g_k}^2\rightarrow \int\text{d}\nu\, \rho_k(\nu)$.

In the following, we make a detailed study for a two-level system (TLA) and a three level atom in a V-configuration (V-system).

 
\subsection{Two-level atom }\label{sec:TLA}

The Hamiltonian in the interaction picture is now
\begin{align}\label{eq:HTLA}
H_I=i\sum_{k=1}^Ng_k(\op{g}{e}a_k^\dagger e^{i\Omega_k t}-\op{e}{g}a_ke^{-i\Omega_kt}),
\end{align}
where $\Omega_k=\nu_k-\omega_{eg}$. The state of the total system and the initial conditions are
\begin{align}
\label{psi_t}
\ket{\Psi(t)}=&(c_g(t)\ket{g}+c_e(t)\ket{e})\ket{0}
+\sum_{k=1}^Nc_k(t)\ket{g}\ket{1_k},\notag\\
c_k(0)=&0,
\end{align}
so that initially the modes of the environment are in a vacuum state. The Schr\"odinger equation and the interaction picture Hamiltonian lead to the following system of first order differential equations for the amplitudes:
\begin{align}\label{eq:TLA_DE}
\dot{c}_g(t)&=0,\notag \\
\dot{c}_e(t)&=-\sum_{k=1}^Ng_ke^{-i\Omega_k t}c_k(t),\notag \\
\dot{c}_k(t)&=g_ke^{i\Omega_kt}c_e(t).
\end{align}
Probabilities to find zero or one photon in the environment are
from Eqs.~\eqref{eq:P_0} and \eqref{eq:P_1} 
\begin{align}
P(0,t)&=\bra{\Psi(t)}\Pi_0\ket{\Psi(t)}=\abs{c_g(t)}^2+\abs{c_e(t)}^2,\notag \\
P(1,t)&=\bra{\Psi(t)}\Pi_1\ket{\Psi(t)}=\sum_{k=1}^N\abs{c_k(t)}^2=1-P(0,t).
\end{align}
In the GAW method, the combined  property states, which are the two possible states 
that the stochastic wave function can take, are by 
using Eq.~\eqref{eq:ModPropertyState}
\begin{align}
\label{prop-S}
\ket{\Psi_0(t)}&=\frac{c_g(t)\ket{g}+c_e(t)\ket{e}}
{\sqrt{\abs{c_g(t)}^2+\abs{c_e(t)}^2}}\ket{0}=\ket{\phi_0(t)} \ket{0},\\ 
\ket{\Psi_{1}(t)}&=\frac{1}{\sum_{j=1}^N\abs{c_k(t)}}
\sum_{k=1}^Nc_k\ket{g}\ket{1_k}\notag\\
&=\frac{1}{\sqrt{P(1,t)}}\sum_{k=1}^Nc_k\ket{g}\ket{1_k}.
\end{align}
Here in the upper equation we use $\ket{\phi_0(t)}$ to denote the reduced system part of the corresponding total system property state. 
From Eqs.~\eqref{eq:J} and~\eqref{eq:Jtot}
we get the combined probability current 
\begin{align}
\J_{1,0}(t)=-2\re{\frac{\dot{c_e}(t)}{c_e(t)}}\abs{c_e(t)}^2.
\end{align}
The probabilities $P(1,t)$ and $P(0,t)$ satisfy
Eq.~\eqref{eq:CombinedGAWMaster} which can be easily calculated 
by using the Hamiltonian and the total state of the 
system and the environment, or the definitions $P(1,t)$, $P(0,t)$ and 
$\J_{1,0}(t)$ [see the text below Eq.\eqref{eq:CombinedGAWMaster}].
We can define the transition rates by using Eqs.~\eqref{eq:GAWTPos} and
\eqref{eq:GAWTNeg}.
When $\J_{1,0}(t)\geq 0$,
\begin{align}\label{eq:GAWTLATPos}
\T_{1,0}(t)&=-2\re{\frac{\dot{c_e}(t)}{c_e(t)}}
\frac{\abs{c_e(t)}^2}{\abs{c_g(t)}^2+\abs{c_e(t)}^2}, \notag\\
\T_{0,1}(t)&=0,
\end{align}
and when $\J_{1,0}(t)< 0$,
\begin{align}\label{eq:GAWTLATNeg}
\T_{1,0}(t)&=0,\notag \\
\T_{0,1}(t)&=2\re{\frac{\dot{c_e}(t)}{c_e(t)}}
\frac{\abs{c_e(t)}^2}{1-\abs{c_g(t)}^2-\abs{c_e(t)}^2}.
\end{align}
Finally, the reduced density matrix can be obtained by taking the trace over the environment
\begin{align*} 
\rho_s(t)&=\tr[\E]{w_0(t)\op{\Psi_0(t)}{\Psi_0(t)}
+w_1(t)\op{\Psi_1(t)}{\Psi_1(t)}}
\notag \\
&=w_0(t)\op{\phi_0(t)}{\phi_0(t)}+w_1(t)\op{g}{g}.
\end{align*}

Next we will study the TLA with the NMQJ method keeping in mind the previously derived results with the GAW method. The master equation for the TLA unraveled with the NMQJ method is 
\begin{align}\label{eq:TLAMaster}
\frac{\text{d}}{\text{d}t}\rho_\Sys(t)&=
-i\left[\frac{1}{2}S(t)\sigma_+\sigma_-,\rho_\Sys(t)\right]\notag \\
&+\Delta(t)
\bigg(\sigma_-\rho_\Sys(t)\sigma_+-
\frac{1}{2}\{\rho_\Sys(t),\sigma_+ \sigma_-\}\bigg),
\end{align}
where the decay rate $\Delta(t)$ and Lamb shift rate $S(t)$ are~\cite{Breuer2007} 
\begin{align}\label{eq:TLARates}
\Delta(t)=&-2\re{\frac{\dot{c}_e(t)}{c_e(t)}},\notag \\ 
S(t)=&-2\im{\frac{\dot{c}_e(t)}{c_e(t)}},
\end{align}
and the non-hermitian Hamiltonian giving the deterministic evolution
of the stochastic wave function is~\cite{Breuer2007}
\begin{align}
H_\eff(t)=\frac{1}{2}\left[S(t)-i\Delta(t)\right]\sigma_+\sigma_-.
\end{align}
All the amplitudes $c_i(t)$ in Eq.~\eqref{psi_t} are solutions of the 
Schr\"odinger equation for the system and the environment with the Hamiltonian
from Eq.~\eqref{eq:HTLA}. These amplitudes have the following connection 
to the normalized state vectors of the effective ensemble of the NMQJ method
\begin{align}
\ket{\psi_0(t)}&=\frac{c_g(t)\ket{g}+c_e(t)\ket{e}}
    {\sqrt{\abs{c_e(t)}^2+\abs{c_g(t)}^2}},
    \notag\\
\ket{\psi_1(t)}&=\ket{g},\notag \\ 
b_g(0)&=c_g(0),\ b_e(0)=c_e(0).
\end{align}
Comparing these with the property state $|\Psi_0(t)\rangle$ of the GAW method in 
Eq.~\eqref{prop-S}, we can see that 
\begin{align}\label{eq:TLAequiv1}
\tr[\E]{\op{\Psi_0(t)}{\Psi_0(t)}}&=\ket{\phi_0(t)} \bra{\phi_0(t)} = \op{\psi_0(t)}{\psi_0(t)}, \notag\\
\tr[\E]{\op{\Psi_1(t)}{\Psi_1(t)}}&= |g\rangle \langle g | = \op{\psi_1(t)}{\psi_1(t)}.
\end{align}
This shows that the reduced system part of the GAW realizations and the NMQJ realizations are identical.
We are left with showing in detail that also the transition rates are the same.

The reduced density matrix in NMQJ is 
\begin{align}\label{eq:NMQJTLArho}
\rho_\Sys(t)&=P(\ket{\psi_0(t)},t)\op{\Psi_0(t)}{\psi_0(t)}\notag \\
&+P(\ket{\psi_1(t)},t)\op{\psi_1(t)}{\psi_1(t)}.
\end{align}
When $\Delta(t)\geq0$ we have transitions from 
$\ket{\psi_0(t)}\rightarrow\ket{\psi_1(t)}$ and from Eq.~\eqref{eq:NMQJRPos}
we obtain
\begin{align}
R_{1,0}(t)=\Delta(t)\frac{\abs{c_e(t)}^2}{\abs{c_g(t)}^2+\abs{c_e(t)}^2}.
\end{align}
This is identical to $\T_{1,0}(t)$, when $\J_{1,0}(t)\geq0$, see 
Eqs.~\eqref{eq:GAWTLATPos} and \eqref{eq:TLARates}.

When $\Delta(t)<0$ we have transitions from 
$\ket{\psi_1(t)}\rightarrow\ket{\psi_0(t)}$ and 
\begin{align}
R_{0,1}(t)=-\frac{P(\ket{\psi_0(t)},t)}{P(\ket{\psi_1(t)},t)}\Delta(t)
\frac{\abs{c_e(t)}^2}{\abs{c_g(t)}^2+\abs{c_e(t)}^2}.
\end{align}
Since $\rho_\Sys(t)$ must be a positive operator we know that the 
decay rate $\Delta(t)$, and therefore also $\J_{1,0}(t)$, must
initially be positive. Let us call $t_1$ the time 
when $\Delta(t)$ turns negative for the  
first time. Now $\rho_\Sys(t)$, when $t<t_1$, generated 
by GAW and NMQJ must be the same, since from Eq. \eqref{eq:TLAequiv1}
we see that the states in the decomposition of $\rho_\Sys(t)$ belong
to the same projective ray 
and $R_{1,0}(t)=\T_{1,0}(t)$ for $t<t_1$.
Therefore we have $P(0,t)={\abs{c_g(t)}^2+\abs{c_e(t)}^2}=P(\ket{\psi_0(t)},t)$ 
and $P(1,t)=P(\ket{\psi_1(t)},t)$. 
Now we can rewrite $R_{0,1}(t)$ as 
\begin{align}\label{eq:TLATisRNeg}
R_{0,1}(t)=&-\frac{P(0,t)}{P(1,t)}\Delta(t)\frac{\abs{c_e(t)}^2}{P(0,t)}\notag\\
=&-\Delta(t)\frac{\abs{c_e(t)}^2}{P(1,t)},
\end{align} 
which is the same
as $\T_{0,1}(t)$, when $\J_{1,0}(t)<0$,
see Eqs.~\eqref{eq:GAWTLATNeg} and~\eqref{eq:TLARates}. It is also clear now that at $t=t_1$ 
both $\J_{1,0}(t)$ and $\Delta(t)$ turn negative.

Thus we have shown that we can derive the NMQJ results from the GAW method for this system.
This means (i) that we can obtain the decay rate in the master equation~\eqref{eq:TLAMaster} 
from the probability currents 
between the total system property states of the GAW method, and (ii) 
that the random state vector in $\Hil_\Sys$ in both methods obtains its possible values from
the same set of states, namely $\ket{g}$ and $\ket{\psi_0}$ (we neglect the global phase since it plays no role here).  

\begin{figure}[ht]
\includegraphics[width=8.6cm]{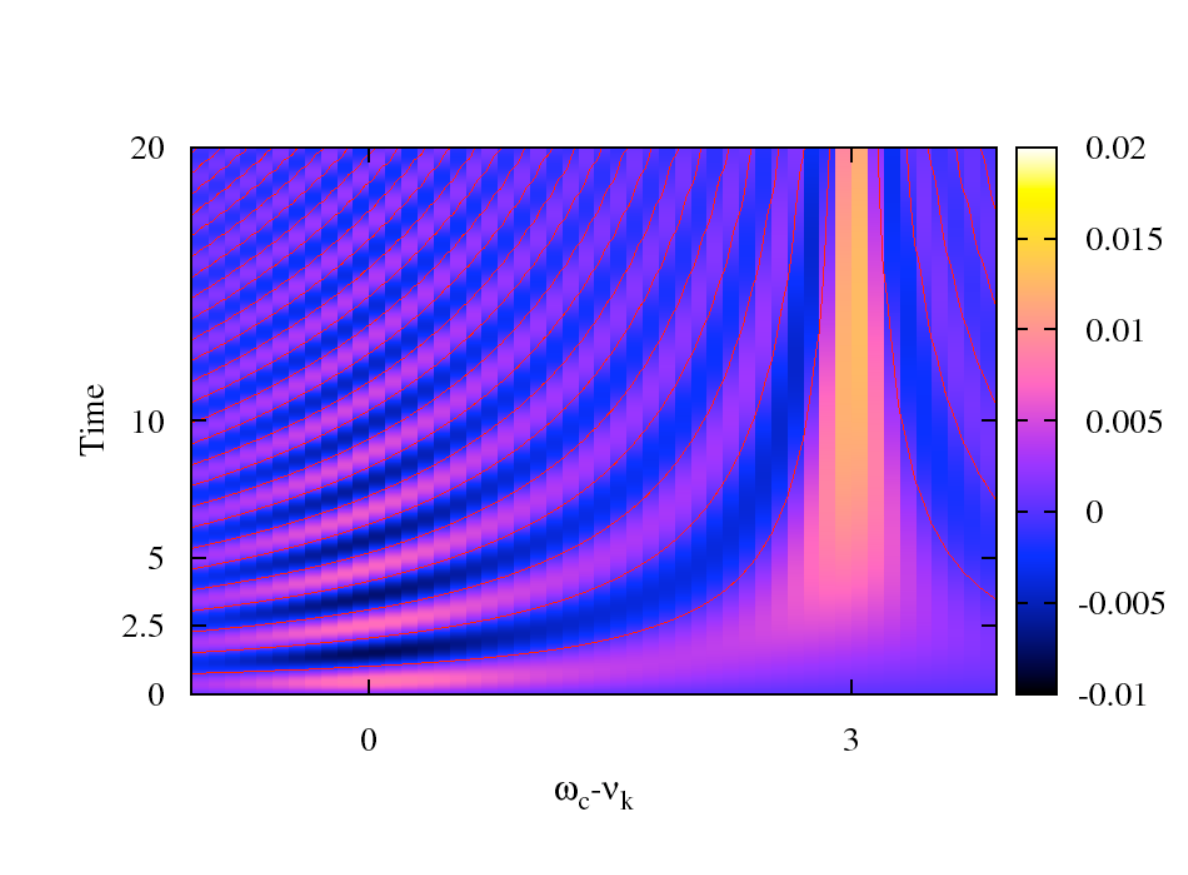}
\caption{\label{Fig:TLAJMarkov} Probability currents $J_{1_k,0}(t)$ for TLA in
the Markovian case. Initial state is $\ket{e}\ket{0}$,
$[t]=1/\lambda$, $\delta=3\lambda$, $\gamma_0=0.8\lambda$ and 
we use 180 environmental modes. 
When $t\approx 1$ we see that there occurs negative probability currents.}
\end{figure}

In the first example we have chosen the parameters
as $\ket{\Psi(t)}=\ket{e}\ket{0}$, time scale $[t]=1/\lambda$, $\delta=3\lambda$ 
and $\gamma_0=0.8\lambda$. We use $180$ environmental modes and a statistical ensemble with $10^4$ members. With the parameters mentioned above, the decay rate in the master 
equation~\eqref{eq:TLAMaster} is time dependent but always positive, thus corresponding to the time-dependent Markovian case~\cite{PhysRevA.81.062115}. 
This also means that there are no reverse jumps in the NMQJ method in this parameter regime. However, as Fig.~\ref{Fig:TLAJMarkov} shows,  
there are negative probability currents in the GAW method for specific modes or individual property states, while the total probability current 
between the system and the environment, $\J_{1,0}(t)$, remains positive indicating net current from the system to the environment. This means that while there are 
individual transitions from one photon to zero photon states in GAW, the number of transitions from zero photon states to one photon states is larger keeping 
the total probability 
current positive, which then matches the probability current obtained from NMQJ.

\begin{figure}[ht]
\includegraphics[width=8.6cm]{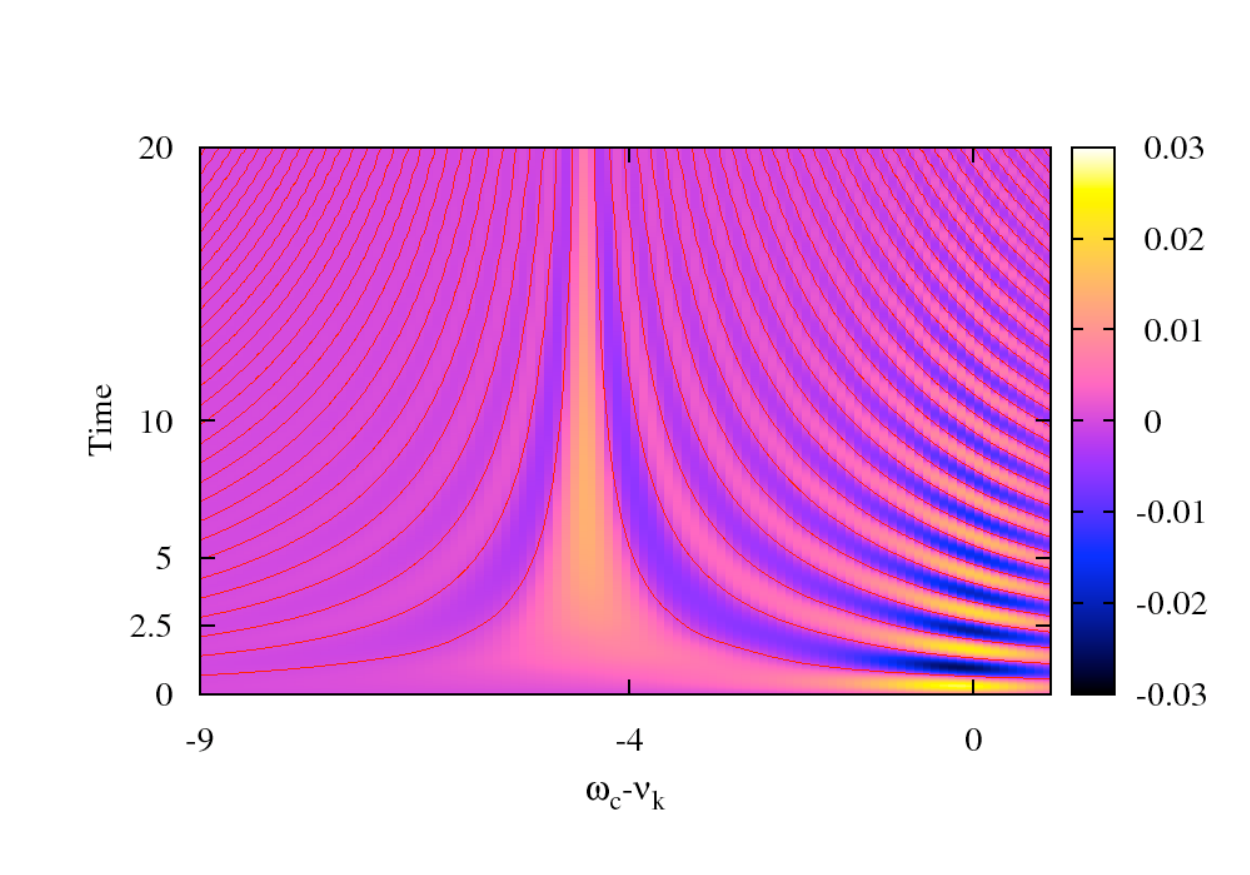}
\caption{\label{Fig:PCurrTLA} Probability currents $J_{1_k,0}(t)$ for TLA. 
The parameters are as in Fig.~\ref{Fig:TLAJMarkov} except that 
$\gamma_0=4\lambda$ and $\delta=-4\lambda$.
We can identify the modes for which $\nu_k\approx\omega_c$ responsible 
for the non-Markovian effects. See the text 
for details.}
\end{figure}

In the second example we have chosen the parameters as 
in the first example except for $\gamma_0=4\lambda$ and $\delta=-4\lambda$.
The system is now in the non-Markovian regime displaying also negative values for the decay rate. 
Figure~\ref{Fig:PCurrTLA} shows the individual probability currents for this case.
The results show that the region $\nu_k-\omega_c \approx 0$ (or $\nu_k \approx \omega_c$) gives the dominant contribution to the total probability current and it has also dominant negative contribution. As a consequence, the total current has negative periods, which is reflected in the negative regions for the decay rate, and thus the system is driven to the non-Markovian regime.

\begin{figure}[ht]
\includegraphics[width=8.6cm]{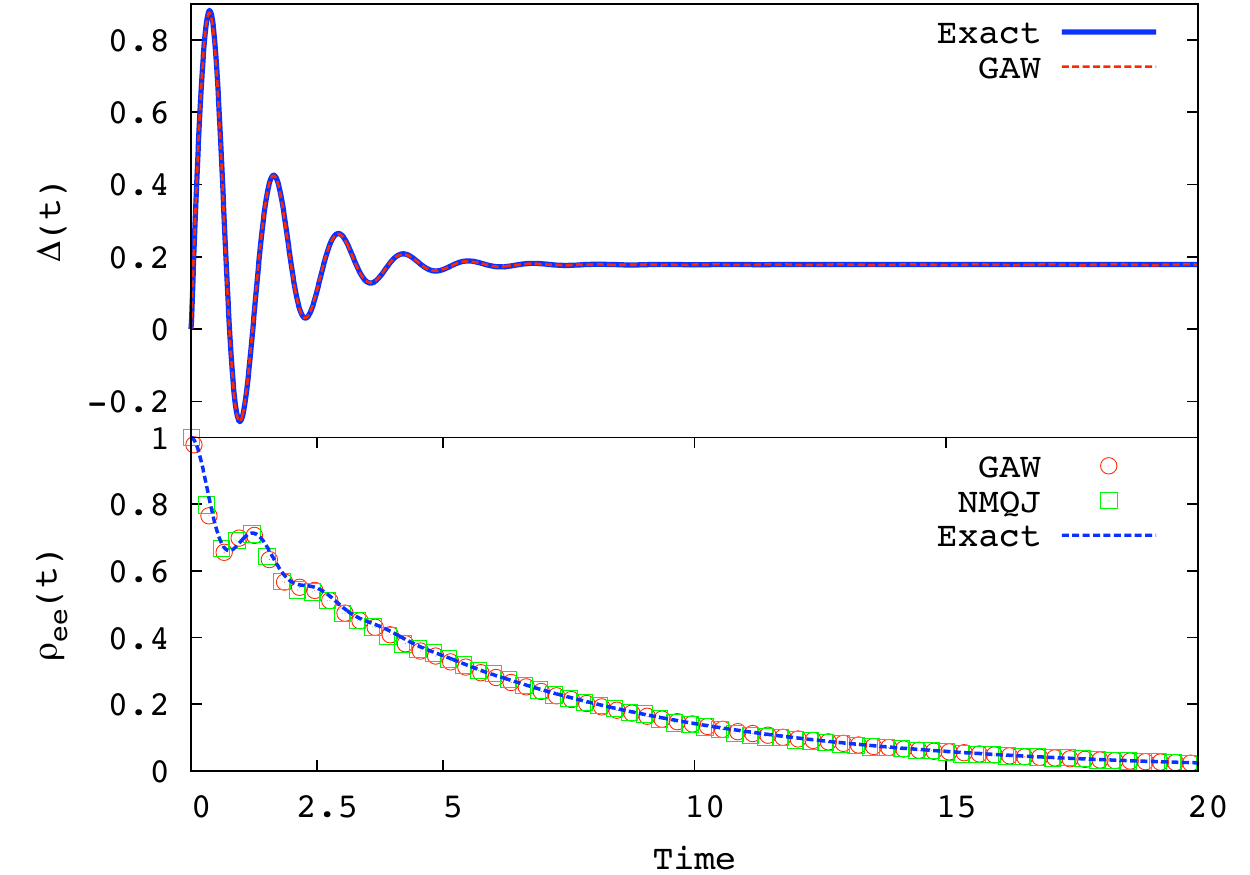}
\caption{\label{Fig:DeltaTLA} Decay rates and excited state 
population for TLA. The parameters are as 
in Fig.~\ref{Fig:PCurrTLA}.} 
\end{figure}

In Fig. \ref{Fig:DeltaTLA} we have plotted the 
decay rate which is calculated from the probability current components.
We compare it to the exact decay rate calculated in the continuum
limit and see that the agreement of the curves is good. In the same figure
we have also plotted the exact solution for the density matrix and 
compare it to the simulated ones, and we can see that the agreement  
of the curves is excellent.

\subsection{V-system}\label{sec:V}

The Hamiltonian for the V-system in the interaction picture is
\begin{align}\label{eq:HVSysNonSec}
H_I=i\sum_{k=1}^Ng_k(\op{c}{a}a_k^\dagger e^{i\Omega_{k,a} t}+
\op{c}{b}a_k^\dagger e^{i\Omega_{k,b} t}+ \text{h.c}),
\end{align}
where $\Omega_{k,i}=\nu_k-\omega_i$, $i={a,b}$, and we have denoted with $|a\rangle$ and $|b\rangle$ the two upper states and with $|c\rangle$ the ground state.
Differential equations for the amplitudes obtained from Schr\"odinger
equation are
\begin{align}\label{eq:VSysDENSec}
\dot{c}_c(t)&=0,\notag \\
\dot{c}_b(t)&=-\sum_{k=1}^Ng_ke^{-i\Omega_{k,b}t}c_k(t) ,\notag \\
\dot{c}_a(t)&=-\sum_{k=1}^Ng_ke^{-i\Omega_{k,a}t}c_k(t) ,\notag \\
\dot{c}_k(t)&=g_k(c_a(t)e^{i\Omega_{k,a}t}+c_b(t)e^{i\Omega_{k,b}t}).
\end{align}
As we have seen in Sec.~\ref{sec:TLA}, the sum of the probability currents 
over all modes is related to the decay rate. By using the same procedure as in 
Sec.~\ref{sec:TLA}, it is possible to derive the  following equations for the probabilities
\begin{align}\label{eq:VSysGAWNSec}
\ddt{t}P(0,t)&=-\J_{1,0}(t),\notag\\
\ddt{t}P(1,t)&=\J_{1,0}(t).
\end{align}
The reduced system dynamics corresponding to these equations is given by
a non-secular master equation which is not in general compatible with the 
form given in Eq.~\eqref{eq:NMMaster} used as a starting point for the NMQJ.

To find the connection between GAW and NMQJ in this system,
we approximate the exact non-secular dynamics by de-coupling the evolution of the coherences and
populations~\cite{Garraway1996}.
Eventually this means that the emission of the photon can be associated to one of the two decay channels
and we write the Hamiltonian as 
\begin{align}\label{eq:HVSysSec}
H_I&=i\sum_{k=1}^Ng_k(\op{c}{a}a_k^\dagger e^{i\Omega_{k,a} t}+
\op{c}{b}b_k^\dagger e^{i\Omega_{k,b} t} + \text{h.c.}),
\end{align} 
where we have introduced new environmental modes described 
by operators $b_k$. This means that we can identify from which decay channel the photon originated, which prevents the occurrence of quantum beats~\cite{PhysRevA.58.440}.

The differential equations for the amplitudes are then
\begin{align}\label{eq:VSysDESec}
\dot{c}_c(t)&=0,\notag \\
\dot{c}_b(t)&=-\sum_{k=1}^Ng_ke^{-i\Omega_{k,b}t}c_k^b(t) ,\notag \\
\dot{c}_a(t)&=-\sum_{k=1}^Ng_ke^{-i\Omega_{k,a}t}c_k^a(t) ,\notag \\
\dot{c}_k^b(t)&=g_kc_b(t)e^{i\Omega_{k,b}t}, \notag \\
\dot{c}_k^a(t)&=g_kc_a(t)e^{i\Omega_{k,a}t},
\end{align}
and these equations are a good approximation for 
Eq.~\eqref{eq:VSysDENSec} in certain parameter regions.

We assume that initially the environmental modes are empty. Then
we can give the total state of the system and the environment as
\begin{align}\label{eq:VSysStateSec}
\ket{\Psi(t)}&=c_c(t)\ket{c}\ket{0}_a\ket{0}_b+c_a(t)\ket{0}_a\ket{0}_b+
c_b(t)\ket{0}_a\ket{0}_b\notag\\
&+ \sum_{k=1}^Nc_k^a(t)\ket{c}\ket{1_k}_a\ket{0}_b+
\sum_{k=1}^Nc_k^b(t)\ket{c}\ket{0}_a\ket{1_k}_b. 
\end{align}
From now on we drop
the subscripts referring to different Hilbert spaces.
We want to know the probabilities to find 
a photon in the environment 
and we want to identify which of the excited states has decayed. 
Therefore it is natural to use the following operators  
\begin{align}\label{eq:VsysProj}
\Pi_0&=I_\Sys\otimes\op{0}{0}\otimes\op{0}{0}, \notag \\
\Pi_{1,a}&=\sum_{k=1}^N I_\Sys\otimes\op{1_k}{1_k}\otimes\op{0}{0}, \notag \\
\Pi_{1,b}&=\sum_{k=1}^N I_\Sys\otimes\op{0}{0}\otimes\op{1_k}{1_k}.
\end{align}
The probability to have one photon in the environment which has been created when the excited state $i$ decayed, is $P^i(1,t)$, where $i=a,b$, and the probability 
to have zero photons in the environment is
$ P(0,t)$. Following a similar procedure as for the earlier presented TLA case, we can calculate them as 
\begin{align}\label{eq:VSysGAWProbs}
P^i(1,t)&=\bra{\Psi(t)}\Pi_{1,i}\ket{\Psi(t)}=\sum_{k=1}^N\abs{c_k^i(t)}^2,\notag \\
P(0,t)&=\bra{\Psi(t)}\Pi_0\ket{\Psi(t)}
=\abs{c_c(t)}^2+\abs{c_b(t)}^2+\abs{c_a(t)}^2. 
\end{align}
Subsequently, the combined property states are
\begin{align}\label{eq:VSysGAWPropertyStates}
\ket{\Psi_0(t)}&=\frac{c_c(t)\ket{c}\ket{0}\ket{0}
  +c_a(t)\ket{a}\ket{0}\ket{0}+c_b(t)\ket{b}\ket{0}\ket{0}}{
\sqrt{P(0,t)}},\\\notag
\ket{\Psi_{1,a}(t)}&=\frac{\sum_{k=1}^Nc_k^a(t)\ket{c}\ket{1_k}\ket{0}}{
\sqrt{P^a(1,t)}},\\\notag
\ket{\Psi_{1,b}(t)}&=\frac{\sum_{k=1}^Nc_k^b(t)\ket{c}\ket{0}\ket{1_k}}{
\sqrt{P^b(1,t)}}.
\end{align}
The differential equations for probabilities $P(0,t)$ and $P^i(1,t)$ 
can be calculated with the help of Eqs.~\eqref{eq:Jtot},~\eqref{eq:VSysDESec} 
and~\eqref{eq:VSysStateSec}.
We obtain
\begin{align}\label{eq:VSysGAWSec}
\ddt{t}P(0,t)&=-\J_{1,0}^a(t)-\J_{1,0}^b(t),\notag\\
\ddt{t}P^a(1,t)&=\J_{1,0}^a(t),\notag\\
\ddt{t}P^b(1,t)&=\J_{1,0}^b(t),
\end{align}
where the combined probability currents $\J_{1,0}^a(t)$ and $\J_{1,0}^b(t)$ tell how much
probability is flowing from the system to the environment in each channel separately. 
The combined probability currents are now
\begin{align}
\J^i_{1,0}(t)=-2\re{\frac{\dot{c_i}(t)}{c_i(t)}}\abs{c_i(t)}^2,
\end{align} 
where $i=a,b$.
We can define transition rates as in 
Eqs.~\eqref{eq:GAWTPos} and~\eqref{eq:GAWTNeg} separately for each
decay path since we can partition the combined probability current 
into two independent parts.
They are, when $\J^i_{1,0}(t)\geq 0$,
\begin{align}\label{eq:VSysGAWTPos}
\T^i_{1,0}(t)&=\frac{\J^i_{1,0}(t)}{P(0,t)}=-2\re{\frac{\dot{c_i}(t)}{c_i(t)}}
\frac{\abs{c_i(t)}^2}{P(0,t)},\notag\\
\T^i_{0,1}(t)&=0,
\end{align} 
and when $\J^i_{1,0}(t)<0$,
\begin{align}\label{eq:VSysGAWTNeg}
\T^i_{1,0}(t)&=0,\notag\\
\T^i_{0,1}(t)&=-\frac{\J^i_{1,0}(t)}{P^i(1,t)}=2\re{\frac{\dot{c_i}(t)}{c_i(t)}}
\frac{\abs{c_i(t)}^2}{P(1,t)},
\end{align}
where $i=a,b$.

The reduced density matrix generated by the GAW method is now
\begin{align}\label{eq:VSysGAWrho}
\rho_\Sys(t)&=\tr[\E]{w_0(t)\op{\Psi_0(t)}{\Psi_0(t)}+w_{1,a}(t)
\op{\Psi_{1,a}(t)}{\Psi_{1,a}(t)}\notag \\
&+w_{1,b}(t)\op{\Psi_{1,b}(t)}{\Psi_{1,b}(t)}}.
\end{align}

Next we will study the NMQJ method for this system.
The master equation
describing the reduced system dynamics under secular approximation is 
\begin{align}\label{eq:VSysMaster}
\frac{\text{d}}{\text{d}t}\rho_\Sys(t)&=
-i[\frac{1}{2}S_a(t)\op{a}{a},\rho_\Sys(t)]
-i[\frac{1}{2}S_b(t)\op{b}{b},\rho_\Sys(t)]\notag \\
&+\Delta_a(t)
\bigg(\op{c}{a}\rho_\Sys(t)\op{a}{c}-
\frac{1}{2}\{\rho_\Sys(t),\op{a}{a}\}\bigg)\notag\\
&+\Delta_b(t)
\bigg(\op{c}{b}\rho_\Sys(t)\op{b}{c}-
\frac{1}{2}\{\rho_\Sys(t),\op{b}{b}\}\bigg).\notag\\
\end{align}
The total state of the system and the environment has been given in
Eq.~\eqref{eq:VSysStateSec}, and by tracing out the environmental
degrees of freedom and taking the time derivative of the 
expression $\op{\Psi(t)}{\Psi(t)}$ we can identify the Lamb shifts and 
the decay rates to be
\begin{align}\label{eq:VsysRates}
\Delta_i(t)&=-2\re{\frac{\dot{c}_i(t)}{c_i(t)}},\\\notag
S_i(t)&=-2\im{\frac{\dot{c}_i(t)}{c_i(t)}},
\end{align}
where $i=a,b$. The non-Hermitian Hamiltonian for NMQJ in this system is 
\begin{align}\label{eq:VsysHeff}
H_\eff(t)=\sum_i\frac{1}{2}(S_i(t)-i\Delta_i(t))\op{i}{i},
\end{align}
where again $i=a,b$.
We can give the deterministically evolving state of the NMQJ process 
and the initial condition as 
\begin{align}\label{eq:VsysNMQJDetState}
\ket{\psi_0(t)}&= d_c(t)\ket{c}+d_a(t)\ket{a}+d_b(t)\ket{b},\\\notag
d_j(0)&=c_j(0),
\end{align}
where $j=c,a,b$ and $c_j(t)$ are probability amplitudes from Eq.~\eqref{eq:VSysStateSec}. 
By solving the time evolution given by the Hamiltonian
of Eq.~\eqref{eq:VsysHeff}, we see that $c_c(t)=d_c(t)$, $c_a(t)=d_a(t)$,
and $c_b(t)=d_b(t)$. In NMQJ, the realizations of the process are normalized and therefore 
we can write 
\begin{align}\label{eq:VSysNMQJStates} 
\ket{\psi_0(t)}&=\frac{c_c(t)\ket{c}+c_a(t)\ket{a}+c_b(t)\ket{b}}{
\sqrt{\abs{c_c(t)}^2+\abs{c_b(t)}^2+\abs{c_a(t)}^2}},\\\notag
\ket{\psi_1(t)}&=\ket{c}
\end{align}
The reduced density matrix of the NMQJ process is then
\begin{align}
\rho_\Sys(t)&=P(\ket{\psi_0(t)},t)\op{\psi_0(t)}{\psi_0(t)}\notag \\
&+P(\ket{\psi_1(t)},t)\op{\psi_1(t)}{\psi_1(t)}.
\end{align}
As in Eqs.~\eqref{eq:NMQJRPos} and~\eqref{eq:NMQJRNeg}, and using 
Eqs.~\eqref{eq:VSysGAWProbs},
we can write the transition rates, when $\Delta_i(t)\geq 0$,
\begin{align}\label{eq:VSysNMQJRPos}
R_{10}^i(t)&=\Delta_i(t)\frac{\abs{c_i(t)}^2}{P(0,t)},\notag \\
R_{01}^i(t)&=0,
\end{align}
and when $\Delta_i(t)<0$,
\begin{align}\label{eq:VSysNMQJRNeg}
R_{10}^i(t)&=0,\notag \\
R_{01}^i(t)&=-\frac{P(\ket{\psi_0(t)},t)}{P(\ket{\psi_1(t)},t)}\Delta_i(t)
\frac{\abs{c_i(t)}}{P(0,t)},
\end{align}
where $i=a,b$. By using Eqs.~\eqref{eq:VSysGAWPropertyStates} and~\eqref{eq:VSysNMQJStates},
we obtain the connection between the GAW and NMQJ state vectors
\begin{align}
\tr[\E]{\op{\Psi_0(t)}{\Psi_0(t)}}&=\op{\psi_0(t)}{\psi_0(t)},\notag \\
\tr[\E]{\op{\Psi_{1,i}(t)}{\Psi_{1,i}(t)}}&=\op{\psi_1(t)}{\psi_1(t)},
\end{align}
where $i=a,b$. 

This means that the system Hilbert space part of the possible
realizations of the combined GAW process and the NMQJ process with the same index belong to the 
the same projective ray in $\Hil_\Sys$. We also see that there is redundancy in
$\tr[\E]{\op{\Psi_{1,i}(t)}{\Psi_{1,i}(t)}}$ since both states when 
$i=a,b$ belong to the same projective ray in $\Hil_\Sys$.
It means that we can combine 
$w_{1,b}(t)+w_{1,a}(t)=w_1(t)$ in Eq.~\eqref{eq:VSysGAWrho} and that 
both transition rates $\T^a$ and $\T^b$ induce jumps 
between the same two projective rays but the 
rates of the jumps are generally different.
 
We assume that from some initial time 
$t_0$ to $t_1$ the rates $\Delta_a(t_0)$ and $\Delta_b(t)$ are positive. Then from
Eqs.~\eqref{eq:VSysGAWTPos},~\eqref{eq:VsysRates} and~\eqref{eq:VSysNMQJRPos} 
we see that 
$R^i_{10}(t)=\T^i_{10}(t)$ and $R^i_{01}(t)=\T^i_{01}(t)$. 
This implies that the density matrices generated
by GAW and NMQJ are the same at least to $t_1$ when at least one of the 
decay rates or collective probability currents turns negative.
Since NMQJ and GAW both 
have normalized realizations we can deduce that 
$w_0(t)=P(0,t)=P(\ket{\psi_0(t)},t)$
and $w_1(t)=P(1,t)=P(\ket{\psi_1(t)},t)$ when $t\in[t_0,t_1]$.

Now when negative currents ($\J^i_{1,0}(t)<0$) or decay rates ($\Delta_i(t)<0$)
emerge when $t>t_1$, the transition  
rates are $\T^i_{0,1}(t)=R^i_{0,1}(t)$ and $\T^i_{1,0}(t)=R^i_{1,0}(t)$ since we 
have $P(0,t)=P(\ket{\psi_0(t)},t)$ and 
$P(1,t)=P(\ket{\psi_1(t)},t)$ (the calculation is the same
as we did in Eq.\eqref{eq:TLATisRNeg}).
Thus we have shown that the GAW process for the combined property state is 
an equivalent process to NMQJ in $\Hil_\Sys$ in the sense we 
defined in the beginning of Sec. \ref{sec:Conn}.

Next we study a numerical example where the initial state is written
in non-secular case as
$\ket{\Psi(t)}=\frac{1}{\sqrt{2}}(\ket{a}_\Sys\ket{0}_\E+\ket{b}_\Sys\ket{0}_\E)$
and under secular approximation, where each channel has independent environment, as
$\ket{\Psi(t)}=\frac{1}{\sqrt{2}}(\ket{a}_\Sys\ket{0}_{\E_a}\ket{0}_{\E_b}+\ket{b}_\Sys\ket{0}_{\E_a}\ket{0}_{\E_b})$.
We are written here the different Hilbert spaces explicitly for clarity but from now on 
we omit this for compactness of notation.
Other parameters are defined as $[t]=1/\lambda$, $\gamma_0=4\lambda$,
 $\delta_a=3\lambda$, $\delta_b=-3\lambda$ and we have used 
$240$ environmental modes.

\begin{figure}[ht]
\includegraphics[width=8.6cm]{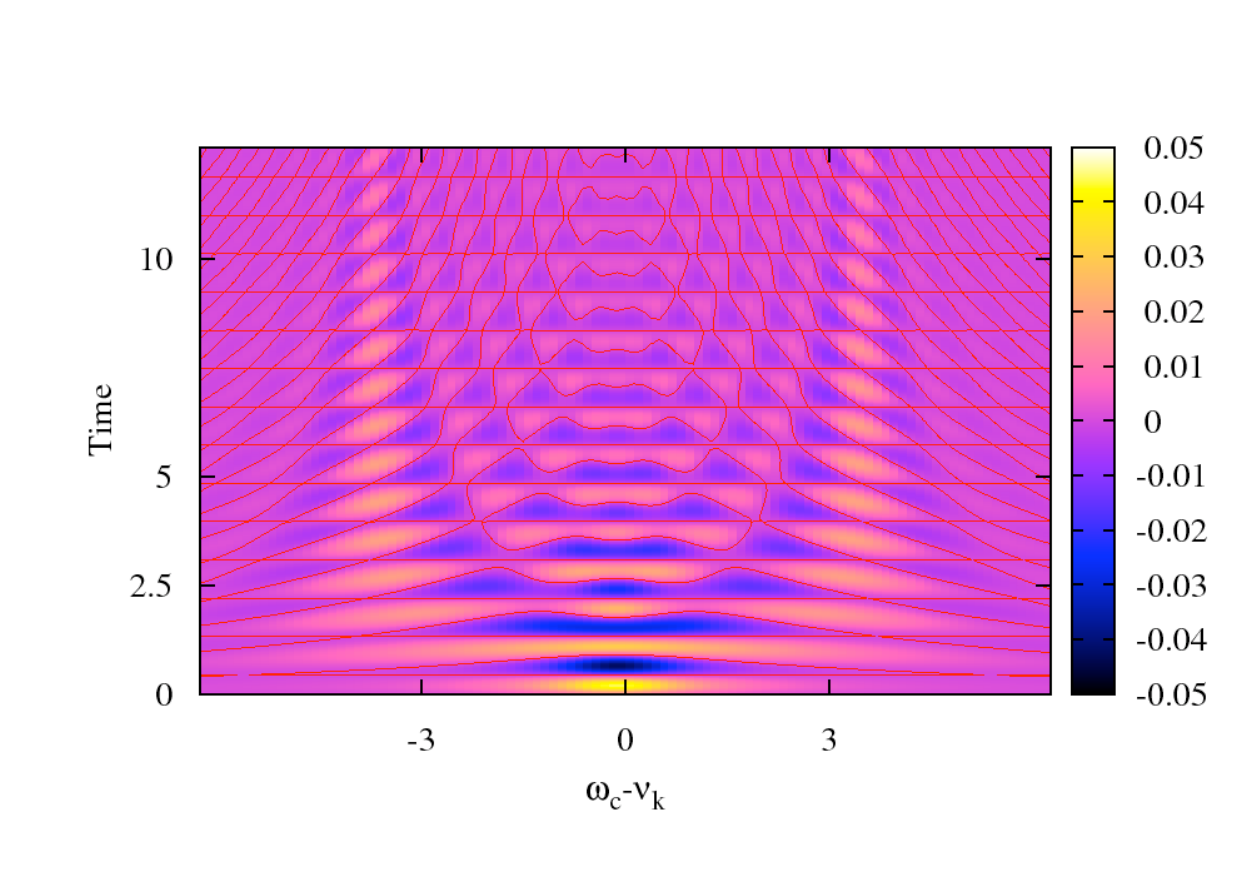}
\caption{\label{Fig:JNonSec} Probability currents 
in the non-secular case. 
The initial state is $\frac{1}{\sqrt{2}}(\ket{a}\ket{0}+\ket{b}\ket{0})$,
$[t]=1/\lambda$, $\gamma_0=4\lambda$,
 $\delta_a=3\lambda$, $\delta_b=-3\lambda$, and we have used 
$240$ environmental modes.}
\end{figure}

\begin{figure}[ht]
\includegraphics[width=8.6cm]{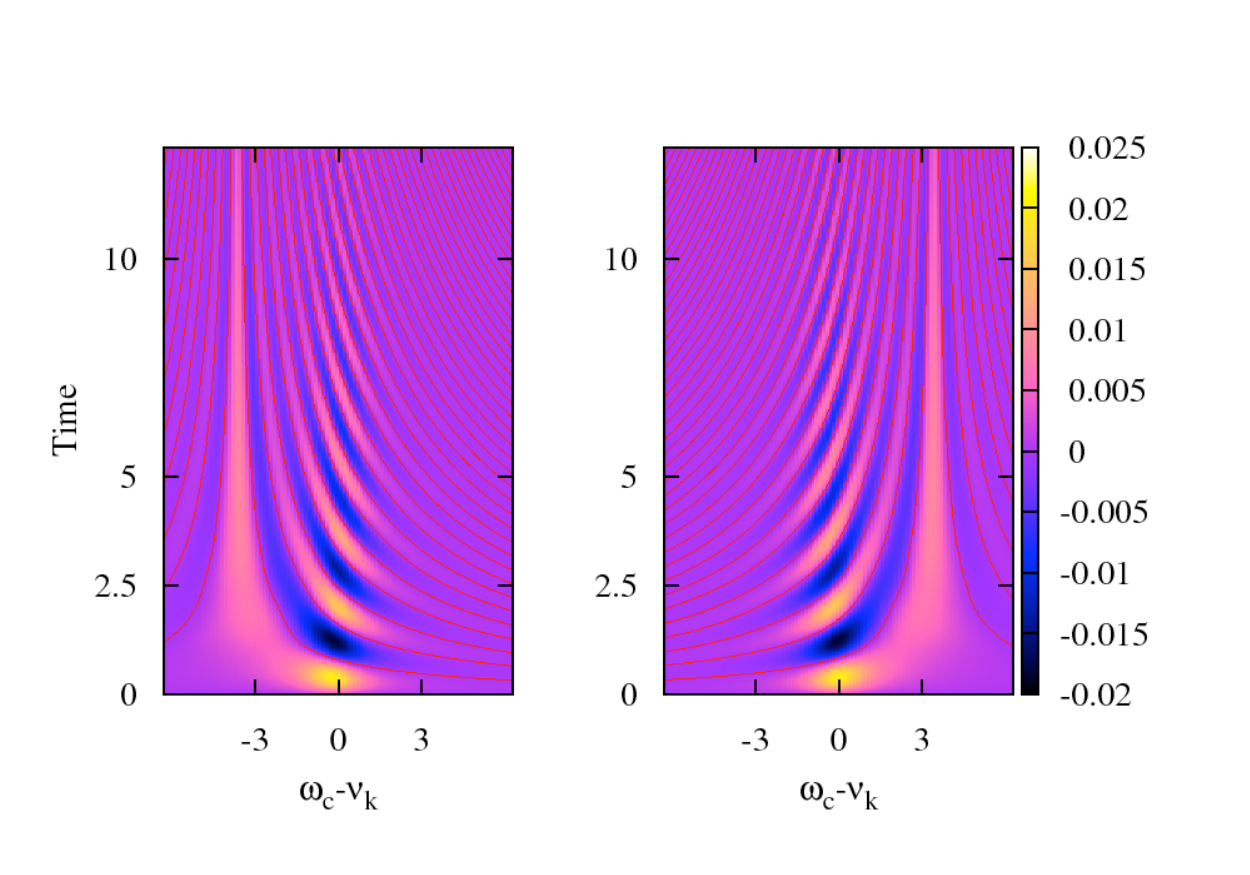}
\caption{\label{Fig:JSec} Probability currents 
in the secular case. The parameters are as in 
Fig.~\ref{Fig:JNonSec} but the initial state is
$\frac{1}{\sqrt{2}}(\ket{a}\ket{0}\ket{0}+\ket{b}\ket{0}\ket{0})$. }
\end{figure}

\begin{figure}
\includegraphics[width=8.6cm]{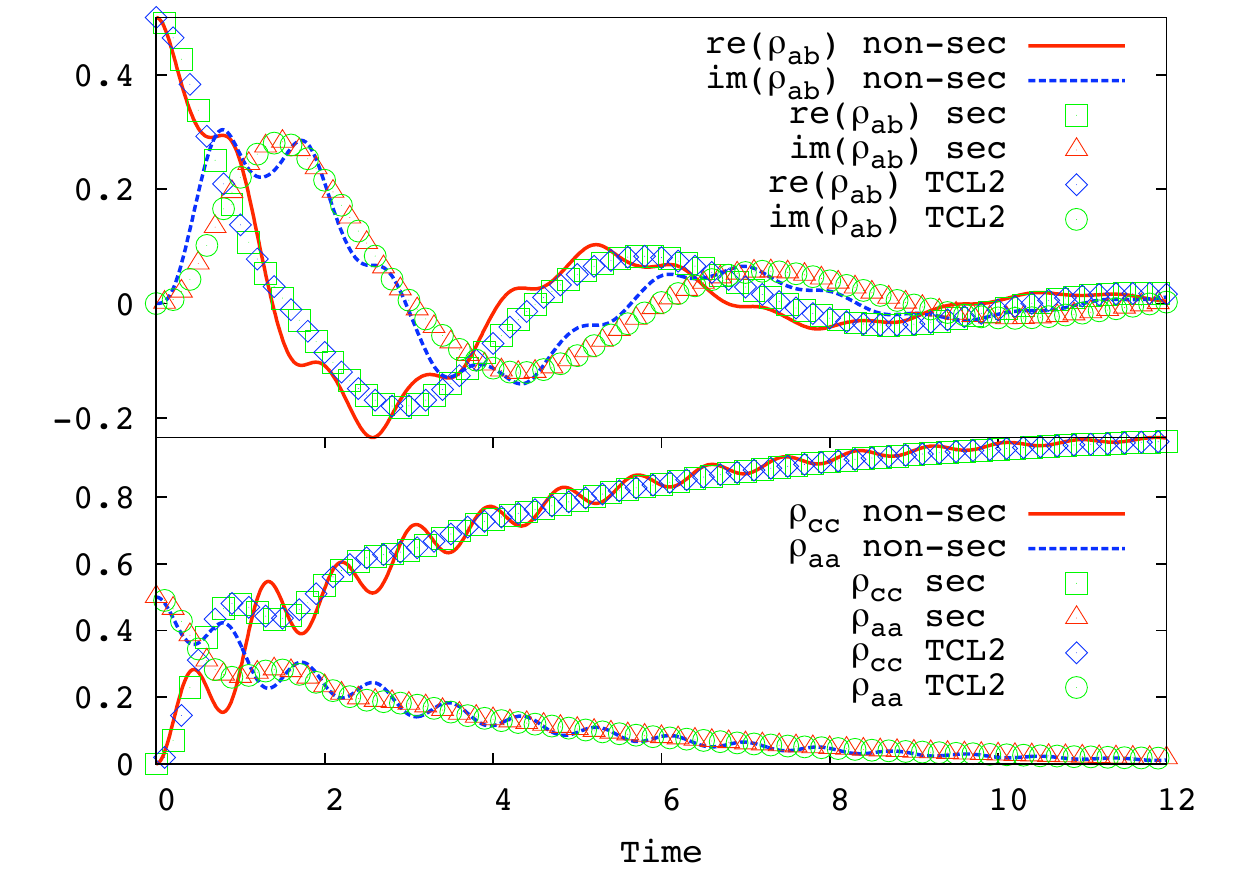}
\caption{\label{Fig:Vsysrho} Density matrix elements for the V-system
The parameters are as in Figs.~\ref{Fig:JNonSec} and~\ref{Fig:JSec}. 
For the chosen parameter values $\rho_{aa}(t)=\rho_{bb}(t)$.}
\end{figure}

We start with the non-secular case. Figure~\ref{Fig:JNonSec} shows the corresponding probability currents between the property states of the GAW method. 
The interference of probability currents is visible. 
In the secular case Fig.~\ref{Fig:JSec} shows probability currents under the secular approximation which decouples the two excited states. Probability currents 
can not interfere because each excited state interacts with its separate environment. 
The comparison between the density matrices for the two cases are shown in Fig.~\ref{Fig:Vsysrho}. In the same figure we can also see that 
under secular approximation the reduced dynamics are governed by 
master equation~\eqref{eq:VSysMaster}. We can clearly see the effect of the interference of the probability currents to the reduced system dynamics in 
the non-secular case.

\begin{figure}[t]
\includegraphics[width=8.6cm]{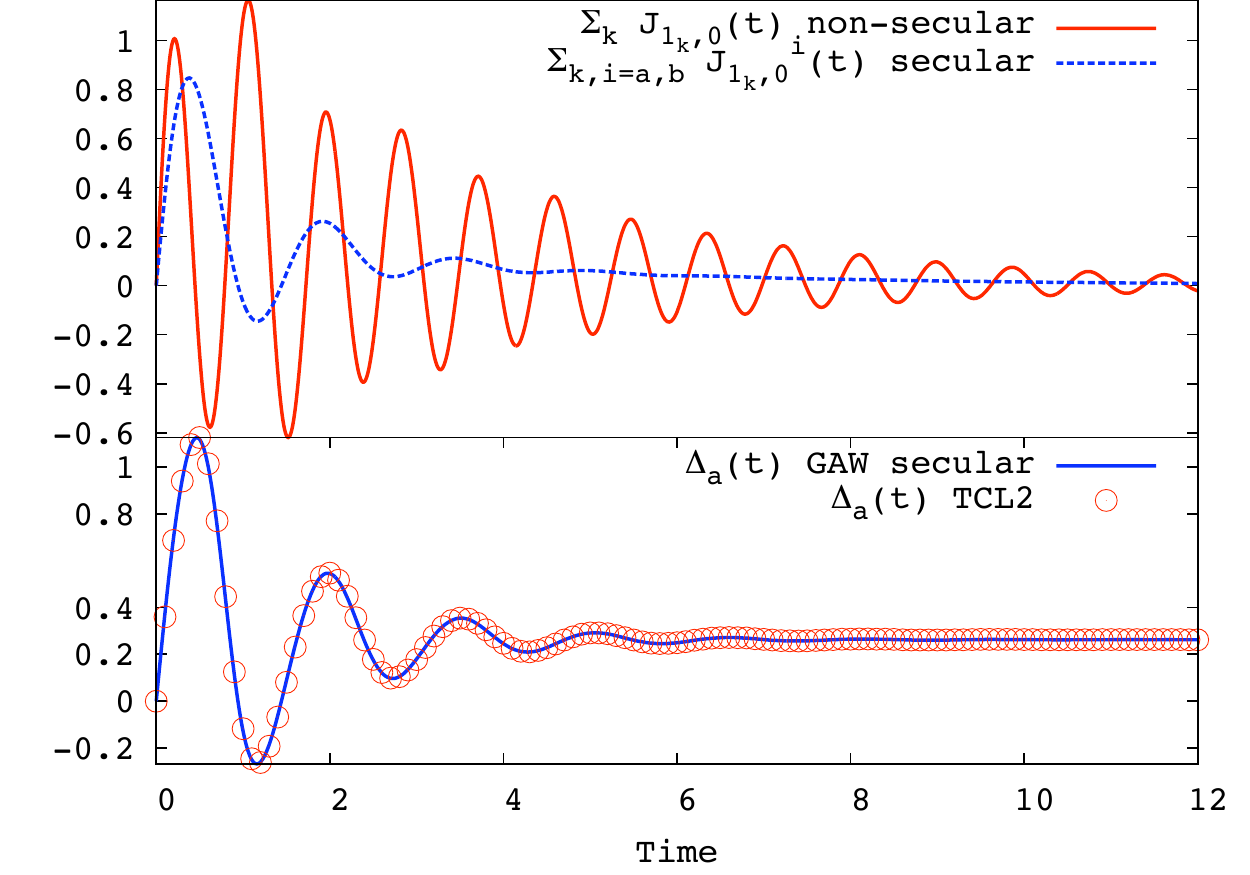}
\caption{\label{Fig:VsysDecAndJ} Top: Sum of probability currents 
from the system to the environment. Solid red line is the non-secular case
where we can not distinguish different decay channels. Blue circles are 
the secular case where we have two different decay channels. 
Bottom: Decay rates calculated from GAW method 
for the secular case (solid blue line) and decay rate of TCL2
master equation (red circles). 
Parameters are as in Figs. \ref{Fig:JNonSec} and \ref{Fig:JSec}.}
\end{figure}

In Fig.~\ref{Fig:VsysDecAndJ} we show the effect of the 
secular approximation to the combined probability current 
from the system to the environment, ie. the difference of
$\J_{1,0}(t)$ and $\J_{1,0}^a(t)+\J_{1,0}^b(t)$. 
There are
fast oscillations in $\J_{1,0}(t)$ and it can even be negative when 
$\J_{1,0}^a(t)+\J_{1,0}^b(t)$ is positive. In Fig.~\ref{Fig:VsysDecAndJ}
we compare the secular approximation decay rate calculated from
GAW to the TCL2~\cite{Breuer2007} decay rate and we see that the match
is very good.

\section{Conclusions}\label{sec:Conc}

We have studied the non-Markovian dynamics of simple quantum optical systems by means of two jumplike unravelings. The GAW method uses piecewise deterministic realizations within the Hilbert space of the total system while the NMQJ method exploits piecewise deterministic realizations within the Hilbert space of the reduced system. Our analysis shows that there exists a connection between the two methods. In particular, we have demonstrated that the reduced system part of the property states of the GAW are identical for the NMQJ state vectors in the considered cases. Moreover, the summation over the probability currents appearing in the GAW formalism are directly connected to the decay rates of the time-local master equations and hence to the rates of jumps in the NMQJ method.
While there exists quite a large variety of Monte Carlo methods for non-Markovian systems
~\cite{PhysRevA.70.012106,PhysRevA.59.1633,BreuerPiilo2009,pseudomodes,PhysRevA.64.053813,Diosi1997569,PhysRevA.69.052104,PhysRevA.66.012108,Garraway1996,PhysRevA.50.3650,PhysRevA.59.2306,Piilo2009,Piilo2008,PhysRevLett.82.1801}, both jump and diffusion type, generally the connections between the methods have not yet been extensively investigated apart of a few studies~\cite{PhysRevA.66.012108,PhysRevA.80.012104}. We expect that the results presented here stimulate further research in this area leading to improved insight to the often complex quantum dynamics of non-Markovian systems. Moreover, analyzing the probability currents in similar manner as treated here, may lead to further understanding of the information flow between the system and the environment, a topic which is currently vividly discussed in the context of open quantum systems. 

\begin{acknowledgments}
This work has been supported by the Academy of Finland (Project No. 133682),
the Magnus Ehrnrooth Foundation and the Vilho, Yrj\"o and Kalle V\"ais\"al\"a Foundation.
We thank J. Gambetta and H.J. Wiseman for stimulating discussions.
\end{acknowledgments}

\bibliography{manu}                                   
                                                           
\end{document}